\documentclass[twocolumn,prb,color,psfig]{revtex4}
\usepackage{graphicx}
\usepackage{epsfig}
\begin{document}
\title{
One-Center Charge Transfer Transitions in Manganites}
\author{A.S. Moskvin}
\affiliation{Department of Theoretical Physics, Ural State University, 620083
Ekaterinburg, Russia}
\date{\today}

\begin{abstract}
In  frames of a rather conventional cluster approach, which combines the
crystal field and the ligand field models
 we have considered different charge transfer (CT) states
and O 2p-Mn 3d CT transitions in MnO$_{6}^{9-}$ octahedra. The many-electron
dipole transition matrix elements were calculated using the Racah algebra for
the cubic point group. Simple "local" approximation allowed to calculate the
relative intensity for all dipole-allowed $\pi -\pi$ and $\sigma -\sigma$ CT
transitions. We present a self-consistent description of the CT bands in
insulating stoichiometric LaMn$^{3+}$O$_3$ compound with the only Mn$^{3+}$
valent state and idealized octahedral MnO$_{6}^{9-}$ centers which allows to
substantially correct the current interpretation of the optical spectra. Our
analysis shows the multi-band structure of the CT optical response
  with the weak low-energy edge at $1.7$ eV, associated with forbidden
$t_{1g}(\pi)-e_{g}$ transition and a series of the  weak and
strong dipole-allowed high-energy transitions starting from $2.5$ and $4.5$ eV,
 respectively, and   extending up to nearly $11$ eV. The most intensive features
 are associated with
 two strong composite bands near $4.6\div 4.7$ eV
and $8\div 9$ eV, respectively,  resulting from the superposition of the dipole-allowed
  $\sigma -\sigma$ and $\pi -\pi$ CT transitions.  These predictions are in
good agreement with experimental spectra. The experimental data point to a
strong overscreening of the crystal field parameter $Dq$ in the CT states of
MnO$_{6}^{9-}$ centers.
\end{abstract}



\maketitle

\section{Introduction}

The discovery of the colossal magnetoresistance in doped manganites like
La$_{1-x}$Sr$_{x}$MnO$_3$ have generated a flurry of the ideas, models and
scenarios
 of this puzzling phenomena, many of which are being developed up to date, although the situation
 remains controversial. There are many thermodynamic and
local microscopic quantities  that cannot be explained by the conventional
double-exchange model with the predominantly Mn 3d location of doped holes. In such
a situation we argue a necessity to discuss all possible candidate states with
different valent structure of  manganese and oxygen atoms, as well as different
valent states of  octahedral MnO$_{6}$ centers. \cite{Avvakumov} Namely these
 slightly distorted
octahedra are believed to be the basic units both for crystalline and
electronic structure in manganites.

 As for an examination of the energy spectrum and
electronic  structure one should emphasize an importance of different optical
 methods because these provide valuable information concerning the dielectric
 function.
The nature of the low-energy optical electron-hole excitations in the
insulating transition metal (3d-) oxides represents one of the most important
challenging issues for these strongly correlated systems. It is now believed
that  the most intensive low-energy electronic excitations
 in insulating 3d-oxides correspond to the transfer of electrons from oxygen anion
to 3d metal cation, hence these materials are  charge transfer (CT) insulators.
 However, the more detailed assignment of these excitations remains open.
All these  are
especially interesting because they could play a central role in multiband Hubbard
models used to describe both the insulating state
and the unconventional states developed under electron or hole doping.

It is now generally accepted that  the ground state of the MnO$_{6}$ centers in
LaMnO$_3$ corresponds to the orbital doublet ${}^{5}E_g$ term of the high-spin
$t_{2g}^{3}e_{g}^1$ configuration. The optical conductivity spectrum of
LaMnO$_3$ exhibits two broad peaks centered around $2.0$ and $5.0$ eV.
\cite{Arima,Okimoto,Jung,Takenaka} However, it has remained unclear just what
the nature of the  intensive
 low-energy optical electron-hole  excitation peaked near $2.0$ eV
 as well as more intensive excitations  with  higher energy peaked
near $5.0$ eV. Some authors \cite{Arima,Okimoto,Takenaka} assign these both
 features  to the dipole-allowed CT transitions like
   $t_{2g}^{3}e_{g}^{1}-t_{2g}^{3}e_{g}^{2}\underline{L}$ and
   $t_{2g}^{3}e_{g}^{1}-t_{2g}^{4}e_{g}^{1}\underline{L}$ ($\underline{L}$
   denoting a ligand hole), respectively. However, others \cite{Jung}
assign the low-energy peak to the "intra-atomic" ${}^{5}E_g-{}^{5}E_{g}^{'}$
transition, or doubly-forbidden (parity and orbital quasimomentum) d-d-like
crystal-field transition between two ${}^{5}E_g$-sublevels separated by a
splitting due to low-symmetry crystalline field.
Both interpretations being particularly qualitative suffer from many
shortcomings and give rise to many questions concerning the details of the
charge transfer states or expected extremely weak intensity for the d-d
transitions.

Unfortunately, the main body of the optical data for manganites is obtained from
reflectivity measurements, that often implies a parasitic contribution due to a
deterioration of the sample surface \cite{Okimoto}, and can give rise to some
ambiguities due to problems
 with Kramers-Kr\"{o}nig transformation.
The more straightforward optical transmission spectra of the LaMnO$_3$ films
 \cite{Lawler} revealed a fine structure of the low-energy $2$ eV band with
 two  features near $1.7$ and $2.4$ eV, respectively, which were
   assigned in contrast with the preceding interpretations
   \cite{Arima,Okimoto,Jung,Takenaka}
  to the Mn$^{3+}$ d-d crystal-field transition ${}^{5}E_{g}-{}^{3}T_{1g}$,
  split by the JT effect.
Such an ambiguity leaves  the question of the nature of the main optical
transitions in LaMnO$_3$ far from being resolved. The band structure
calculations, including the LDA+U, fail to clear up the situation because of
these cannot reproduce the important effects of intra-atomic correlations
forming the term structure both of   ground and excited  CT states. In this
connection a rather conventional quantum-chemical cluster approach, which
combines the crystal field and the ligand field models with real opportunity to
include all correlation effects, seems more relevant.

Below, in  frames of such an approach we consider different CT states and O
2p-Mn 3d CT transitions in MnO$_6$ octahedra.  As we suppose,  they define the
main part of the optical response both for undoped LaMn$O_3$ and different
 doped manganites.

\section{Electronic structure of manganese ions and manganese-oxygen
octahedral centers in manganites}
 Five manganese Mn 3d and eighteen  oxygen O 2p atomic
orbitals in octahedral MnO$_6$ complex with  the point symmetry group $O_h$
form both hybrid Mn 3d-O 2p  bonding and antibonding $e_g$ and $t_{2g}$
molecular orbitals (MO), and non-bonding $a_{1g}(\sigma)$, $t_{1g}(\pi)$,
$t_{1u}(\sigma)$, $t_{1u}(\pi)$, $t_{2u}(\pi)$ ones.
\cite{Ber,Douglas,TMO,Kahn} Non-bonding $t_{1u}(\sigma)$ and $t_{1u}(\pi)$ with
the same symmetry  are hybridized due to the oxygen-oxygen O 2p$\pi$ - O
2p$\pi$ transfer. The relative energy position of different non-bonding oxygen
orbitals is of primary importance for the spectroscopy of the oxygen-manganese
charge transfer. This is firstly determined by the bare energy separation
$\Delta \epsilon _{2p\pi \sigma}=\epsilon _{2p\pi }-\epsilon _{2p\sigma}$
between O 2p$\pi$ and O 2p$\sigma$ electrons.
\begin{figure}[t]
\includegraphics[width=8.5cm,angle=0]{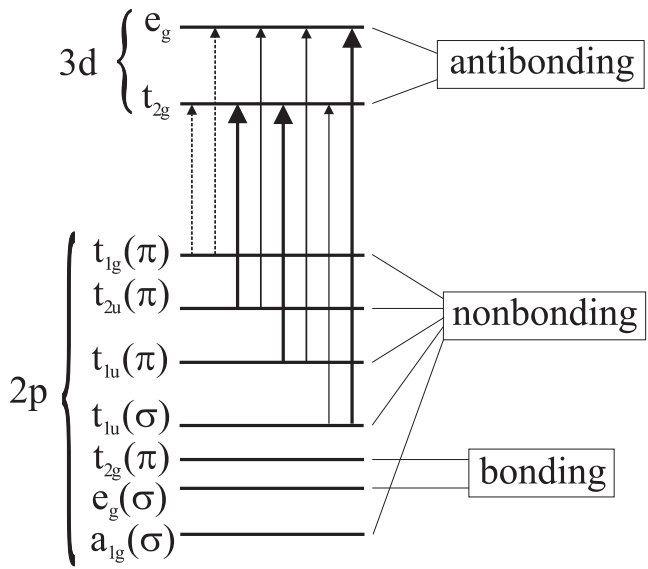}
\caption{The diagram of Mn 3d-O 2p molecular orbitals  for the MnO$_{6}$
octahedral center. The O 2p - Mn 3d charge transfer transitions are shown by
arrows: strong dipole-allowed  $\sigma -\sigma$  and $\pi -\pi$ by thick solid
arrows;  weak dipole-allowed  $\pi -\sigma$ and $\sigma -\pi$ by thin solid
arrows; weak dipole-forbidden low-energy transitions by thin dashed arrows,
respectively.} \label{fig1}
\end{figure}
  Since the O 2p$\sigma$ orbital points towards the two neighboring positive
   $3d$ ions, an electron in this orbital has its energy lowered
   by the Madelung potential as compared with the O 2p$\pi$ orbitals,
   which are perpendicular
   to the respective 3d-O-3d axes. Thus, Coulomb arguments  favor
    the positive sign of the $\pi -\sigma$ separation
     $\epsilon _{p\pi}-\epsilon _{p\sigma}$ which
  numerical value   can be easily
   estimated in frames of the well-known point charge model \cite{Ber}, and appears to be of the order of
   $1.0$ eV.
   In a first approximation, all the $\gamma (\pi )$ states
    $t_{1g}(\pi),t_{1u}(\pi),t_{2u}(\pi)$ have the same energy. However,
the O 2p$\pi$-O 2p$\pi$ transfer yields the energy correction
   to its bare energies with the largest in value
   and positive in sign  for
   the $t_{1g}(\pi)$ state. The energy of $t_{1u}(\pi)$ state drops due to
   hybridization with cation 4p$t_{1u}(\pi)$ state. In other words, the $t_{1g}(\pi)$ state is believed
   to be the highest in energy non-bonding oxygen state.
   For illustration, in Fig.1 we show the energy spectrum of the 3d-2p manifold in
   octahedral complexes like MnO$_6$ with the relative energy position of the levels
   according to the $X_{\alpha}$-method calculations \cite{Licht} for the
   FeO$_{6}^{9-}$ octahedral complex in a lattice environment typical for perovskites
   like LaFeO$_3$, LaMnO$_3$.

 The {\it conventional} electronic structure of octahedral MnO$_6$
complexes is associated with the configuration of the completely filled O 2p shells
and partly filled Mn 3d shells. The typical high-spin ground state
configuration and crystalline term for Mn$^{3+}$ in octahedral crystal field or
for the octahedral MnO$_{6}^{9-}$ center is $t_{2g}^{3}e_{g}^1$ and
${}^{5}E_{g}$, respectively. Namely this orbital doublet results in a vibronic
coupling and Jahn-Teller (JT) effect for the MnO$_{6}^{9-}$ centers, and
cooperative JT ordering in LaMnO$_3$. In the framework of crystal field model
the ${}^{5}E_{g}$ term originates from the  (3d$^{4}\,{}^{5}D)$ term of  free
Mn$^{3+}$ ion. Among the low-energy crystal field d-d transitions for the
high-spin Mn$^{3+}$ ions one should note a single spin-allowed and
parity-forbidden ${}^{5}E_{g}-{}^{5}T_{2g}$ transition at energy varying from
about $2.0$  to $2.5$ eV depending on the crystalline matrix. So, this is
likely to be near $2.5$ eV for the Mn$^{3+}$ impurity in perovskite YAlO$_3$.
\cite{Gen}
 The transition is magneto-optically active, and could manifest itself in the Faraday and
  Kerr effects. A detection of the spin- and parity-forbidden
  ${}^{5}E_{g}-{}^{3}T_{1g}$
  transition with, probably, lower energy could be rather important in  the Mn$^{3+}$
  assignment.

The {\it unconventional} electronic configuration of octahedral MnO$_6$
complexes is associated with a {\it charge transfer  state} with  one hole in O
2p shells. The excited CT configuration $\underline{\gamma}_{2p}^1$ 3d$^{n+1}
\;$ arises from the transition of an electron from the MO predominantly anionic
in nature (the $\gamma_{2p}$ hole  in the core of the anionic MO being hereby
produced), into an empty 3d-type MO ($t_{2g}\,$ or $\,e_g$). The transition
between the ground configuration and the excited one can be presented as the
$\,\gamma_{2p}\,\rightarrow\,$ 3d$(t_{2g},e_g)$ CT transition.

The CT configuration consists of two partly filled subshells, the ligand $\;
\gamma _{2p}\,$-, and the cation 3d$(t_{2g}^{n_{1}}e_{g}^{n_{2}})$ shell,
respectively. It should be emphasized that  the oxygen hole having occupied the
{\it non-bonding} $\; \gamma _{2p}\,$ orbital interact {\it ferromagnetically}
with 3d$(t_{2g}^{n_{1}}e_{g}^{n_{2}})$ shell. This rather strong 
ferromagnetic coupling results in Hund rule for the CT configurations, and
provides the high-spin ground states. The maximal value of the total spin for
the Hund-like CT state in MnO$_{6}^{9-}$ center equals $S=3$, that uncovers
some perspectives to unconventional magnetic signatures of these states.
\cite{Avvakumov}


It should be noted that the presence of the oxygen hole moving around 3d-ion in
the CT state can, in common, provide a strong screening both of the 3d crystal
field and intra-atomic electron-electron repulsion with the renormalization of
the appropriate  correlation Racah parameters $A,B,C$ and crystal field
splitting $Dq$. It should  be noted that unlike the "monopole" parameter $A$
the "multipole" ones $B$ and $C$ usually manifest relatively weak response to
crystalline environment. \cite{Ber} Nevertheless, all these effects can
strongly complicate the calculation of the energy structure for the respective
3d$^{n+1}$ configuration. This configuration in the case of CT states in
MnO$_{6}^{9-}$ center nominally corresponds to Mn$^{2+}$ ion.
\begin{figure}[h]
\includegraphics[width=8.5cm,angle=0]{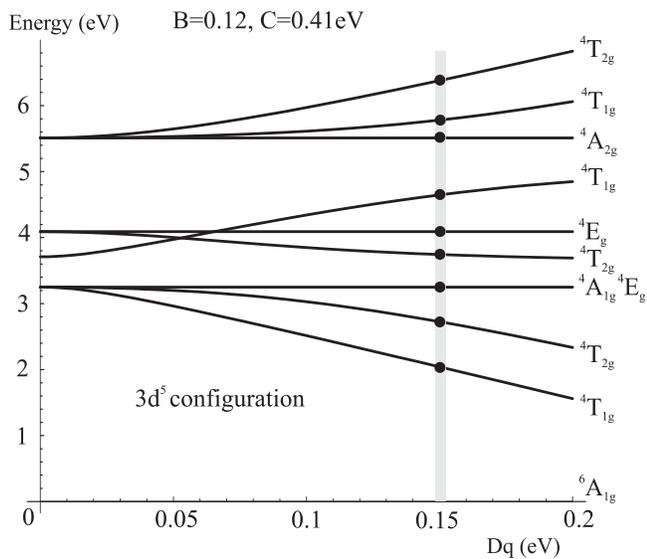}
\caption{The energy-level Tanabe-Sugano diagram for the high-spin terms of 3d$^5$
configuration in octahedral (or cubic) crystal field.} \label{fig2}
\end{figure}
In Fig.2 we reproduce the conventional octahedral crystal field energy
 scheme (Tanabe-Sugano diagram) for the
 high-spin states of 3d$^{5}$ configuration given the parameters
  $B=0.12$, $C=0.41$ eV typical for Mn$^{2+}$ ions in oxides. \cite{Orgel,Elp}
It is well-known that Mn$^{2+}$ ion in cubic crystal field or  the octahedral
MnO$_{6}^{10-}$ center manifests usually the high-spin ground state
${}^{6}A_{1g}$ of the $t_{2g}^{3}e_{g}^2$ configuration.
  The strong Coulomb repulsion leading to the high-spin ${}^{6}A_{1g}$
 ground  term for Mn$^{2+}$ ion would result in strong corrections to a simple
 picture of the energies for the CT states based on the one-electron approach
 sketched in Fig.1. Indeed, the one-electron model predicts the lower energy
 of the  CT states with 3d$^{5};t_{2g}^{4}e_{g}^1$ configuration rather than
 3d$^{5};t_{2g}^{3}e_{g}^2$ configuration. This example clearly demonstrates
 the role played by intra-atomic correlations.
   It should be noted that for the 3d$^{5}$ configuration the
  ${}^{6}A_{1g}$ term is the only spin-sextet, so all the d-d transitions from the
   ground state are spin- and parity-forbidden, therefore these are extremely
    weak (oscillator strength $\sim 10^{-7}$) \cite{MnO}, and hardly observable in the optical
    absorption spectra. However, it has been recently shown
    (see Ref. \onlinecite{MnO} and references therein)
     that the dipole- and spin-forbidden d-d excitations can be examined excellently by
    spin-polarized electron energy loss spectroscopy (EELS). Fromme {\it et al}.  \cite{MnO}
have measured with high accuracy ($\leq 0.02$ eV) the excitation energies for
almost all quartet ${}^{4}\Gamma$ terms of Mn$^{2+}$ ions in MnO:
  ${}^{4}T_{1g}({}^{4}G)$ ($2.13$), ${}^{4}T_{2g}({}^{4}G)$ ($2.4$),
  ${}^{4}A_{1g},{}^{4}E_{g}({}^{4}G)$ ($2.82$),
  ${}^{4}T_{2g}({}^{4}D)$ ($3.31$), ${}^{4}E_{g}({}^{4}D)$ ($3.82$),
   ${}^{4}T_{1g}({}^{4}P)$ ($4.57$), ${}^{4}A_{2g}({}^{4}F)$ ($5.08$),
${}^{4}T_{1g}({}^{4}F)$ ($5.38$ eV). The energy of ${}^{4}T_{2g}({}^{4}F)$
term is expected to be near $6.0$ eV. \cite{Orgel} Simple textbook
 three-parameter ($B, C, Dq$) crystal field theory \cite{Tanabe}
  provides a self-consistent description of these data given the crystal
   field parameter $Dq= 0.15$ eV (see Fig.2). Rigorously speaking, we should
   conclude $Dq= \pm 0.15$ eV because the energies of all quartet and sextet
   terms for the half-filled 3d$^5$ configuration depend only on
    modulus $|Dq|$. Parameter $Dq$ defines not only the value of the cubic
    crystal field splitting $\Delta = 10 Dq = E(e_{g})-E(t_{2g})$, but its sign.
In other words, measuring only the energy of quartet terms for Mn$^{2+}$ (or
Fe$^{3+}$) we cannot separate two opposite cases: $E(e_{g})>E(t_{2g})$ and
$E(e_{g})<E(t_{2g})$. However, this puzzling effect concerns only the energy
rather than wave functions for the ${}^{4}T_g$ terms which  strongly depend on
the sign of $Dq$. Given  $Dq > 0$ the lowest ${}^{4}T_{1g}$ term is of dominant
$t_{2g}^{4}e_{g}^{1}$ configuration, while for $Dq < 0$ this is of dominant
$t_{2g}^{2}e_{g}^{3}$ configuration. Below we shall see that the latter effect
would result in very strong difference in the lineshape of the CT bands for $Dq
> 0$ and  $Dq < 0$ given the same energy spectrum of the CT states.

Finally, one should emphasize that both the energy and wave functions for
 ${}^{6}A_{1g},{}^{4}A_{1g},{}^{4}A_{2g},{}^{4}E_{g}$ terms do not depend
 on the crystal field parameter $Dq$ at all.

\section{Charge transfer transitions in
M\lowercase{n}O$_{6}^{9-}$ centers}

A set of the intensive and broad absorption bands in parent manganites is
usually assigned to the anion-cation O 2p-Mn 3d charge transfer.
\cite{Arima,Okimoto,Takenaka} In the framework of the MnO$_{6}$ center model
this elementary CT process generates both intra- and inter-center CT
transitions for the MnO$_{6}^{9-}$ centers. The intra-center CT transitions
could be associated with the {\it small CT  Frenkel excitons}
 \cite{Davydov} and represent the oxygen hole moving around 3d$^{n+1}$-cation.
  The inter-center CT transitions form a set of {\it small CT  excitons},
  which  in terms of chemical notions represent somewhat like the {\it disproportionation}
  quanta
\begin{equation}
  MnO_6^{9-}+MnO_6^{9-} \rightarrow MnO_6^{10-}+MnO_6^{8-}
  \label{dispro}
\end{equation}
resulting in a formation of the bounded electron MnO$_6^{10-}$  ($e$-) and hole
MnO$_6^{8-}$ ($h$-) small radius centers.  A minimal energy of such an exciton
or the disproportionation
   threshold usually proves to be lower than appropriate purely
   ionic quantity which value equals to the electrostatic correlation energy $U_{dd}\approx
   10$ eV. The estimates made in Ref.\onlinecite{Kovaleva} yield for this
   energy in LaMnO$_3$: $\Delta E \approx 3.7$ eV.

\subsection{Intra-center charge transfer transitions in  MnO$_{6}^{9-}$ centers}

The conventional classification scheme of the CT transitions in the octahedral
MnO$_{6}^{9-}$ centers (intra-center CT transitions) incorporates the
electric-dipole allowed transitions $\gamma _{u} \rightarrow$ 3d$t_{2g},3de_g$
from the odd oxygen
 $\gamma _{u} =t_{1u}(\pi),t_{2u}(\pi),t_{1u}(\sigma)$ orbitals to the even
  manganese 3d$t_{2g}$ and 3d$e_g$  orbitals, respectively.
  These one-electron   transitions
   generate a manifold of the many-electron
  ones ${}^{5}E_{g}\rightarrow {}^{5}T_{u}$ ($T_u$ may be equal both to
  $T_{1u}$, and $T_{2u}$) which may additionally
  differ by the crystalline term of the respective 3d$^{n+1}$ configuration:
\begin{equation}
(t_{2g}^{3}{}^{4}A_{2g};e_{g}^1){}^{5}E_{g}\rightarrow
((t_{2g}^{4}{}^{3}T_{1g};e_{g}^{1}){}^{4}\Gamma_{g};
\underline{\gamma _{u}}){}^{5}T_{u},
\end{equation}
\begin{equation}
(t_{2g}^{3}{}^{4}A_{2g};e_{g}^1){}^{5}E_{g}\rightarrow
((t_{2g}^{3}{}^{4}A_{2g};e_{g}^{2};{}^{2S_{1}+1}\Gamma_{1g}){}^{2S+1}\Gamma_{g};
\underline{\gamma _{u}}){}^{5}T_{u}
\end{equation}
for $\gamma _{u} \rightarrow$ 3d$t_{2g}$ and $\gamma _{u} \rightarrow$ 3d$e_g$
transitions, respectively. Here, we already took into account the Pauli
principle and the "triangle rules" for spin momenta and orbital "quasi-momenta"
like $\Gamma$. Proceeding with this analysis we can obtain full selection rules
for the many-electron CT  transitions:

1. Each  $\gamma _{u} \rightarrow$ 3d$t_{2g}$ transition generates two doublets
of many-electron CT  transitions ${}^{5}E_{g}\rightarrow {}^{5}T_{1,2u}$,
differing by the term of the $t_{2g}^{4}e_{g}^{1}$ configuration:
${}^{4}T_{1g}$ and ${}^{4}T_{2g}$, respectively.

2. Each $\gamma _{u} \rightarrow$ 3d$e_{g}$ transition generates 5
many-electron CT  transitions ${}^{5}E_{g}\rightarrow {}^{5}T_{1,2u}$,
differing by the term of the $t_{2g}^{3}e_{g}^{2}$ configuration:
${}^{4}A_{1g}$, ${}^{4}A_{2g}$, ${}^{6}A_{1g}$, ${}^{4}E_{g}$, respectively (in
the case of ${}^{4}E_{g}$ term each one-electron transition generates the
${}^{5}T_{1,2u}$ doublet). The $\Gamma _{1g}$ quasi-momentum is determined by
the triangle rule: $\Gamma _{1g} = A_{2g}\otimes \Gamma _{g}$.

Additionally, we should take into account  the configurational interaction.
Indeed, the terms with the same symmetry for different configurations interact
and mix with each other. In our case, for the 3d$^5$ configuration
 we have 3 terms ${}^{4}T_{1g}$ and 3 terms ${}^{4}T_{2g}$
 which present both in
the $t_{2g}^{4}e_{g}^{1}$ and $t_{2g}^{3}e_{g}^{2}$, $t_{2g}^{2}e_{g}^{3}$
configurations (see Fig.2). For the $t_{2g}^{3}e_{g}^{2}$ configuration there
are two interacting ${}^{4}E_{g}$ terms.

Hence, beginning from 3 $t_{1u}(\pi),t_{1u}(\sigma),t_{2u}(\pi)$ non-bonding
purely oxygen orbitals as initial states for one-electron CT we come to 60 (!)
many-electron dipole-allowed CT  transitions ${}^{5}E_{g}\rightarrow
{}^{5}T_{1,2u}$: 24 transitions $t_{1u}(\pi),t_{2u}(\pi)-t_{2g}$ ($\pi -\pi$
channel), 16 transitions $t_{1u}(\pi),t_{2u}(\pi)-e_{g}$ ($\pi -\sigma$
channel), 12 transitions $t_{1u}(\sigma)-t_{2g}$ ($\sigma -\pi$ channel), and 8
transitions $t_{1u}(\sigma)-e_{g}$ ($\sigma -\sigma$ channel), respectively.
Thus, the real situation with the multi-band structure of the dipole-allowed O
2p - Mn 3d CT transitions in MnO$_{6}^{9-}$ centers  requires the crucial
revision of the oversimplified approaches based on the concept of the only  O
2p - Mn 3d CT transition \cite{Kovaleva} or at the best of two O 2p - Mn
3d$t_{2g},e_{g}$ CT transitions. \cite{Arima,Okimoto,Takenaka}

\subsection{Dipole transition matrix elements}
Making use of the Racah algebra (the method of the irreducible tensorial
operators) \cite{Var,Tanabe,Bostrem} both for spins and quasimomenta we obtain
after some cumbersome calculations the following expressions for the dipole
transition matrix elements between the ground
$(t_{2g}^{3}{}^{4}A_{2g};e_{g}^1){}^{5}E_{g}$ and excited CT states in
MnO$_{6}^{9-}$ octahedra:
\begin{widetext}
{\bf  $\gamma _{u}\rightarrow t_{2g}$ transfer}

$$
\langle (t_{2g}^{3}{}^{4}A_{2g};e_{g}^1){}^{5}E_{g}\mu|{\hat d}_{q}|
((t_{2g}^{4}{}^{3}T_{1g};e_{g}^{1}){}^{4}\Gamma_{g};
\underline{\gamma _{u}}){}^{5}T_{u}\mu ^{'}\rangle
=
$$
\begin{equation}
(-1)^{\mu}
 \left\langle \begin{array}{ccc}
E_{g} & T_{1u} & T_{u} \\
-\mu & q & \mu ^{'}
\end{array}
\right\rangle ^{*}
\langle (t_{2g}^{3}{}^{4}A_{2g};e_{g}^1){}^{5}E_{g}
\|{\hat d}\|((t_{2g}^{4}{}^{3}T_{1g};e_{g}^{1}){}^{4}\Gamma_{g};
\underline{\gamma _{u}}){}^{5}T_{u}\rangle \,,
\label{t}
\end{equation}
where for the many-electron submatrix element
$$
\langle (t_{2g}^{3}{}^{4}A_{2g};e_{g}^1){}^{5}E_{g}
\|{\hat d}\|((t_{2g}^{4}{}^{3}T_{1g};e_{g}^{1}){}^{4}\Gamma_{g};
\underline{\gamma _{u}}){}^{5}T_{u}\rangle =
$$
\begin{equation}
(-1)^{j(\gamma _{u})}\, 3\sqrt{2}
[\Gamma _{g}]
\left\{ \begin{array}{ccc}
T_{2} & A_{2} & T_{1} \\
E & \Gamma  & E
\end{array}
\right\}
\left\{ \begin{array}{ccc}
\gamma & T_{1} & T_{2} \\
E & \Gamma  & T
\end{array}
\right\}
\langle \gamma _{u}\|d\|t_{2g}\rangle \, :
\label{t1}
\end{equation}

{\bf $\gamma _{u}\rightarrow e_{g}$ transfer}

$$
\langle (t_{2g}^{3}{}^{4}A_{2g};e_{g}^1){}^{5}E_{g}\mu|{\hat d}_{q}|
((t_{2g}^{3}{}^{4}A_{2g};e_{g}^{2};{}^{2S_{1}+1}\Gamma_{1g})
{}^{2S+1}\Gamma_{g};\underline{\gamma _{u}}){}^{5}T_{u}\mu ^{'}\rangle
=
$$
\begin{equation}
(-1)^{\mu}
\left\langle \begin{array}{ccc}
E_{g} & T_{1u} & T_{u} \\
-\mu & q & \mu ^{'}
\end{array}
\right\rangle ^{*}
\langle (t_{2g}^{3}{}^{4}A_{2g};e_{g}^1){}^{5}E_{g}\|{\hat d}\|
((t_{2g}^{3}{}^{4}A_{2g};e_{g}^{2};{}^{2S_{1}+1}\Gamma_{1g})
{}^{2S+1}\Gamma_{g};\underline{\gamma _{u}}){}^{5}T_{u}\rangle \,,
\label{e}
\end{equation}
where for the many-electron submatrix element
$$
\langle (t_{2g}^{3}{}^{4}A_{2g};e_{g}^1){}^{5}E_{g}\|{\hat d}\|
((t_{2g}^{3}{}^{4}A_{2g};e_{g}^{2};{}^{2S_{1}+1}\Gamma_{1g})
{}^{2S+1}\Gamma_{g};\underline{\gamma _{u}}){}^{5}T_{u}\rangle =
$$
\begin{equation}
(-1)^{1+j(\gamma _{u})+j(\Gamma )}\, \sqrt{6}
[S_{1},S,\Gamma _{1g},\Gamma _{g}]^{1/2}
\left\{ \begin{array}{ccc}
3/2 & S_{1} & S \\
1/2 & 2 & 1/2
\end{array}
\right\}
\left\{ \begin{array}{ccc}
A_{2} & E & E \\
E & \Gamma  & \Gamma _{1}
\end{array}
\right\}
\left\{ \begin{array}{ccc}
\gamma & T_{1} & E \\
E & \Gamma  & T
\end{array}
\right\}
\langle \gamma _{u}\|d\|e_{g}\rangle \, .
\label{e1}
\end{equation}
\end{widetext}
Here, the expressions (\ref{t}), (\ref{e}) represent the Wigner-Eckart theorem
for many-electron transition matrix elements with $\left\langle
\begin{array}{ccc} \cdot & \cdot & \cdot \\ \cdot & \cdot & \cdot
\end{array}
\right\rangle $ being the Wigner coefficient for the cubic point group.
 \cite{Tanabe,Bostrem}
In (\ref{t1}), (\ref{e1}) the conventional notations are used for spin $6j$-
and orbital $6\Gamma$ symbols ($\left\{ \begin{array}{ccc} \cdot & \cdot &
\cdot \\ \cdot & \cdot & \cdot
\end{array}
\right\} $); $[S]=2S+1$, $[\Gamma]$ is the dimensionality of the corresponding
irreducible representation of the cubic point group; $j(\Gamma )$ the so-called
quasimomentum number; $\langle \gamma _{u}\|\hat d\|\gamma _{g}\rangle$ is the
one-electron dipole moment submatrix element. The latter is defined by the
respective  Wigner-Eckart theorem as follows
\begin{equation}
\langle \gamma _{u}\mu|{\hat d}_{q}|\gamma _{g}\mu ^{'}\rangle =
(-1)^{j(\gamma _{u})-\mu}
\left\langle \begin{array}{ccc}
\gamma _{u} & T_{1u} & \gamma _{g} \\
-\mu & q & \mu ^{'}
\end{array}
\right\rangle ^{*} \langle \gamma _{u}\|\hat d\|\gamma _{g}\rangle \,.
\end{equation}
It should be noted that the above calculation of the dipole matrix elements
implies the same character of the one-electron $t_{2g},e_g$ orbitals both for
the ground state $t_{2g}^3,e_{g}^1$ and excited $t_{2g}^3,e_{g}^2$
$t_{2g}^4,e_{g}^1$ CT configurations, particularly  with the invariable values
of the appropriate covalency parameters.

The one-electron dipole moment submatrix elements can be rather simply
evaluated in frames of the so-called "local" approximation, in which  the
calculation of the matrix of the dipole moment implies the full neglect all
many-center integrals:
$$
\langle \phi _{k_{1}}({\bf R}_{1}-{\bf r})|\hat{\bf d}| \phi _{k_{2}}({\bf
R}_{2}-{\bf r})\rangle = e{\bf R}_{1} \delta _{{\bf R}_{1},{\bf R}_{2}}\,\delta
_{k_{1},k_{2}}\, ,
$$
where ${\bf R}_{1},{\bf R}_{2}$ label sites, $k_{1},k_{2}$ atomic functions,
respectively. Then
$$
\langle t_{2u}(\pi )\|\hat d\|e_{g}\rangle =0;\, \langle t_{2u}(\pi )\|\hat
d\|t_{2g}\rangle = -i\sqrt{\frac{3}{2}}\lambda _{\pi}d \,;
$$
\begin{equation}
\langle t_{1u}(\sigma )\|\hat d\|t_{2g}\rangle =0;\, \langle t_{1u}(\sigma
)\|\hat d\|e_{g}\rangle = -\frac{2}{\sqrt{3}}\lambda _{\sigma}d \, ;\label{d}
\end{equation}
$$
\langle t_{1u}(\pi )\|\hat d\|e_{g}\rangle =0;\, \langle t_{1u}(\pi )\|\hat
d\|t_{2g}\rangle = \sqrt{\frac{3}{2}}\lambda _{\pi}d\, .
$$
Here, $\lambda _{\sigma}, \lambda _{\pi}$ are $effective$ covalency parameters
 for $e_{g},t_{2g}$ electrons, respectively, $d=eR_0$ is an elementary dipole moment
 for the cation-anion bond length $R_0$.
 We see, that the "local" approximation results in an additional selection rule:
 it forbids the $\sigma \rightarrow \pi$, and  $\pi \rightarrow \sigma $
 transitions, $t_{1u}(\sigma )\rightarrow t_{2g}$, and
 $t_{1,2u}(\pi )\rightarrow e_{g}$, respectively, though these are dipole-allowed.
In other words, in frames of this approximation only $\sigma$-type
($t_{1u}(\sigma )\rightarrow e_{g}$) or $\pi$-type ($t_{1,2u}(\pi )\rightarrow
t_{2g}$) CT transitions are allowed. It should be emphasized that the "local"
approximation, if non-zero, provides the leading contribution to transition
matrix elements with corrections being of the first order in the cation-anion
overlap integral. Interestingly, that the one-electron dipole moment submatrix
elements for both $\pi \rightarrow \pi $  transitions in Eq. (\ref{d}) have the
same absolute value. Hereafter, we make use of the terminology of "strong" and
"weak" transitions for the dipole-allowed CT transitions going on the $\sigma
-\sigma$, $\pi -\pi$, and $\pi -\sigma$, $\sigma -\pi$ channels, respectively.
 Thus, for MnO$_{6}^{9-}$
center we predict a series of 32 strong many-electron
dipole-allowed CT  transitions ${}^{5}E_{g}\rightarrow {}^{5}T_{1,2u}$
(24 for $\pi -\pi$, and 8 for $\sigma -\sigma$ channel) and 28 weak
dipole-allowed CT transitions
($\pi -\sigma$ and  $\sigma -\pi$ cross-channels).

 The formulas (\ref{t1})-(\ref{d}) together with the numerical values of some
 $6j$- and $6\Gamma$-symbols,
 listed below in Appendix allow to make quantitative predictions for the relative magnitude
 of the intensities for different CT transitions.
 First of all we would like to compare the overall integral intensities for the
 strong dipole-allowed CT transitions in the $\pi -\pi$ and  $\sigma -\sigma$ channels.
  To this end, we calculate and sum
the line strengths (the dipole submatrix element squared) which are
proportional to the appropriate oscillator strengths:
\begin{equation}
I_{\pi \pi}=9 \lambda _{\pi}^{2}\,d^{2};\,\,
I_{\sigma \sigma }=\frac{3}{2} \lambda _{\sigma }^{2}\,d^{2}\,,
\end{equation}
or
\begin{equation}
I_{\pi \pi}/ I_{\sigma \sigma }=6 \lambda _{\pi}^{2}/\lambda _{\sigma }^{2}\,.
\end{equation}
In other words, the ratio of the total oscillator strengths for these channels
is determined by the ratio of the respective cation-anion charge density
transfer parameters. Usually, $\lambda _{\sigma }^{2}>\lambda _{\pi}^{2}$,
however, it seems the overall intensity for the $\pi -\pi$  channel can be
comparable with that of for $\sigma -\sigma$ channel, or even exceed it. In
frames of the separate channel we can obtain exact relations between the
partial oscillator strengths for the CT transitions differing by the final
state of the $t_{2g}^{4}e_{g}^1$ and
 $t_{2g}^{3}e_{g}^2$  configuration for the
 $\pi -\pi$ and $\sigma -\sigma$ channel, respectively.
For the $\pi -\pi$ channel we have
\begin{equation}
I({}^{4}T_{1g};{}^{5}T_{u}):I({}^{4}T_{2g};{}^{5}T_{u})=
\frac{1}{2}\,:\,\frac{1}{2}
\label{I1}
\end{equation}
irrespective of the type of the transferred oxygen $\pi$ electron, $t_{2u}$, or
$t_{1u}$. Here, each intensity represents the sum for two doublets,
${}^{5}T_{1u}$ and ${}^{5}T_{2u}$. Interestingly, the relative intensity for
these two components is $3\,:\,1$ and $1\,:\,3$ for the ${}^{4}T_{1g}$ and
${}^{4}T_{2g}$ "intermediate" terms, respectively. It should be noted that
absolutely the same relations are fulfilled for the weak dipole-allowed CT
transitions in the $\sigma -\pi$  channel.

For the strong $\sigma -\sigma$, and weak $\pi -\sigma$ channels we have the
more nontrivial relation:
$$
I({}^{6}A_{1g};{}^{5}T_{u}):I({}^{4}A_{1g};{}^{5}T_{u}) :
 I({}^{4}E_{g};{}^{5}T_{u}): I({}^{4}A_{2g};{}^{5}T_{u})
$$
\begin{equation}
=\frac{8}{15}\,:\,\frac{2}{15}\,:\,\frac{1}{9}\,:\,\frac{2}{9} \, , \label{I2}
\end{equation}
where the third transition is doublet with the equal intensity of both
components. It should be reminded that all these numerical data are obtained
for CT transitions to "pure" $t_{2g}^{4}e_{g}^1$  and
 $t_{2g}^{3}e_{g}^2$  configurations. The configuration interaction effect
 results in a redistribution of the respective intensities in between all
 interacting terms with the same symmetry.

\section{Charge transfer transitions in parent L\lowercase{a}M\lowercase{n}O$_3$}
Now we can apply the model theory to the undoped stoichiometric manganite
LaMnO$_3$. For our analysis to be more quantitative we make two rather obvious
model approximations. First of all, one assumes that for MnO$_{6}^{9-}$ centers
in LaMnO$_3$ as usually for cation-anion octahedra in  3d-oxides
\cite{TMO,Kahn,Licht} the non-bonding $t_{1g}(\pi)$ oxygen orbital has the
highest energy and forms the first electron removal oxygen state. Moreover, to
be definite we assume that the energy spectrum of the non-bonding oxygen states
for Mn$^{3+}$O$_{6}^{9-}$ centers in LaMnO$_3$
 coincides with that calculated in Ref.\onlinecite{Licht}
for Fe$^{3+}$O$_{6}^{9-}$ in LaFeO$_3$ with the same crystalline environment,
 in other words, we have (in eV):
 $$
 \Delta (t_{1g}(\pi)-t_{2u}(\pi))\approx 0.8\, ;\,\,
 \Delta (t_{1g}(\pi)-t_{1u}(\pi))\approx 1.8\, ;
 $$
 $$
 \Delta (t_{1g}(\pi)-t_{1u}(\sigma))\approx 3.0\, .
 $$
This  is believed to be a rather reasonable choice of the energy parameters,
because the purely oxygen states mainly depend only on crystalline environment.
Secondly, we choose for the Racah parameters $B$ and $C$ the numerical values
typical for Mn$^{2+}$ in oxides, $0.12$ and $0.41$ eV, respectively (see
above). The crystal-field parameter $Dq$ may be varied, however, we decide in
favor  of only two model considerations with the same absolute value but the
different sign of $|Dq|=0.15$ eV ("conventional" $Dq>0$  and "unconventional"
$Dq<0$ sign). Let us mention that this value, irrespective of the sign,
provides a reasonable explanation of the Mn$^{2+}$ spectra  in MnO \cite{MnO}
(see above). Moreover, the photoemission data  \cite{Park} are believed to
confirm the relevance of this value for crystal field splitting parameter in
LaMnO$_3$.

Hereafter, this set of parameters is used for the model theoretical simulation
of the overall CT band in LaMnO$_3$. Firstly, we argue that the lowest in
energy spectral feature observed in LaMnO$_3$
 near $1.7$ eV is
believed to be associated with the onset of  the series of the dipole-forbidden
CT transitions $t_{1g}(\pi)\rightarrow e_{g},t_{2g}$, rather than with any d-d
crystal field transition. The energy of this transition was picked out to be a
starting point to assign all other CT transitions.

Weak dipole-allowed $\pi -\sigma$ CT transitions $t_{2u}(\pi)-e_{g}$ and
$t_{1u}(\pi)-e_{g}$ form more intensive CT bands starting at higher than the
preceding series energies,  near $2.5$  and $3.5$ eV, respectively, in
accordance with the magnitude of the $t_{1g}(\pi)-t_{2u}(\pi)$ and
$t_{1g}(\pi)-t_{1u}(\pi)$ separations. Actually, the $t_{1u}(\pi)-e_{g}$
transition has to be more intensive because the $t_{1u}(\pi)$ state is partly
hybridized with $t_{1u}(\sigma)$, hence this transition borrows a portion of
intensity from the strong dipole-allowed $t_{1u}(\sigma )-e_g$ CT transition.

 The latter $\sigma -\sigma$ transition as we see
from Eq.(\ref{I2}) forms intensive broad CT band starting from the  main
${}^{5}E_{g}- {}^{6}A_{1g};{}^{5}T_{1u}$ peak at $\approx 4.7$ eV and ranging
to the  ${}^{5}E_{g}- {}^{4}A_{2g};{}^{5}T_{2u}$
 peak at $\approx 10.2$ eV with interstitial peaks at $\approx 8.0$ eV being the
 result of the superposition of two transitions
 ${}^{5}E_{g}- {}^{4}A_{1g};{}^{5}T_{1u}$ and
 ${}^{5}E_{g}- {}^{4}E_{g};{}^{5}T_{u}$, and at $\approx 8.8$ eV
 due to another ${}^{5}E_{g}- {}^{4}E_{g};{}^{5}T_{u}$ transition,
 respectively.
 Thus, the overall width of the CT bands with final
 $t_{2g}^{3}e_{g}^2$  configuration occupies a spectral range from
 $1.7$  up to
 $\sim 10$ eV.

 As it is seen from Eq.(\ref{I1}) strong dipole-allowed $\pi -\pi$ CT transitions
  $t_{2u}(\pi),t_{1u}(\pi)-
 t_{2g}$ form two manifolds of equally intensive CT bands shifted with respect each other
 by the $t_{2u}(\pi)-t_{1u}(\pi)$ separation ($\approx 1.0$ eV).
 In turn, each manifold consists of two triplets  of weakly split and
 equally intensive CT bands
 associated with ${}^{5}E_{g}- {}^{4}T_{1g};{}^{5}T_{u}$ and
 ${}^{5}E_{g}- {}^{4}T_{2g};{}^{5}T_{u}$ transitions, respectively.
 In accordance with the   assignment of crystal-field transitions
  \cite{MnO}  in LaMnO$_3$ (see Fig.2) we should expect the low-energy
  edge of the
 dipole-allowed $\pi -\pi$ CT band starting from $\approx 4.5$ eV
 ($1.7+2.0+$($t_{1g}(\pi)-t_{2u}(\pi)$ separation)). Taking account of strong
 configuration interaction we should expect the high-energy edge of this band
 associated with the highest in energy ${}^{4}T_{2g}$ term of the 3d$^5$
 configuration to be situated near $\approx 9.9$ eV. In between, in accordance
  with our scheme of energy levels we predict peaks at
 5.2; 5.5; 6.2 ($\times 2$);  7.2($\times 2$); 7.9; 8.2; 8.3; 8.9 eV.
 The weak dipole-allowed $\sigma -\pi$ transitions occupy the high-energy
 spectral range from  $6.7$ to $11.1$ eV.
\begin{figure*}[t]
\includegraphics[width=18.0cm,angle=0]{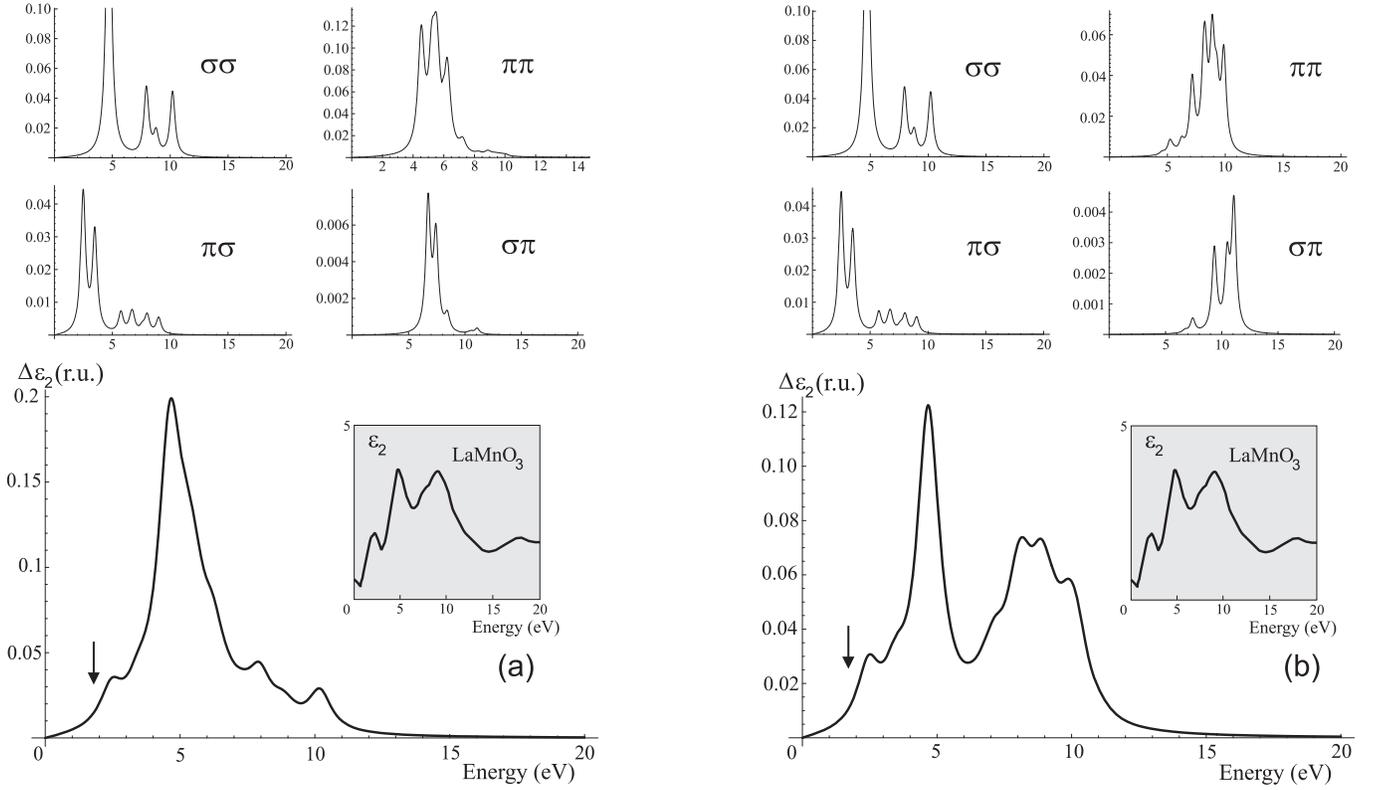}
\caption{Theoretical simulation of the overall O 2p-Mn 3d CT band in LaMnO$_3$
with conventional $Dq>0$ (a) and unconventional $Dq<0$ (b) sign  for the
crystal field parameter. The top panels shows the partial contributions of
different dipole-allowed transitions. The lower panels present the overall
contribution of the dipole-allowed CT transitions to the imaginary part of
dielectric function. Experimental spectrum for La$_{1-x}$Sr$_x$MnO$_3$ given
$x\approx 0$ from the paper by Okimoto {\it et al.}\cite{Okimoto} is shown in
insert (see text for details).} \label{fig3}
\end{figure*}

 Overall, our analysis shows  the multi-band structure of the CT optical response
 in LaMnO$_3$ with the weak low-energy edge at $1.7$ eV, associated with forbidden
$t_{1g}(\pi)-e_{g}$ transition and a series of  strong bands in the range
$4.6\div 10.2$ eV beginning from composite peak at $\sim 4.5\div 4.7$ eV
and  closing by composite peak at $8\div 10$ eV both resulting from
the superposition of strong dipole-allowed
$\pi -\pi$ and $\sigma -\sigma$ CT transitions.

Above we addressed the model energies of the CT transitions. In frames of our
model approach the relative intensities for different dipole-allowed CT
transitions are governed  by the relative magnitude of different one-electron
dipole submatrix elements. We have performed the theoretical simulation of the
overall O 2p-Mn 3d CT optical band  in  LaMnO$_{3}$ generated by dipole-allowed
CT transitions in MnO$_{6}^{9-}$ octahedra given simple physically reasonable
assumptions concerning the one-electron submatrix elements. We have assumed: i)
the equal integral intensities for the $\sigma -\sigma$ and $\pi -\pi$
channels: $I_{\sigma \sigma}=I_{\pi \pi}$, that corresponds $\lambda
_{\sigma}^2 =6\lambda _{\pi}^2$; and ii)  the equal integral intensities
 $I_{\pi \sigma}=I_{\sigma \pi}=0.1I_{\sigma \sigma}$  for all weak
dipole-allowed transitions $t_{2u}(\pi)-e_g$, $t_{1u}(\pi)-e_g$, and
$t_{1u}(\sigma)-t_{2g}$. The former assumption agrees with the well-known
semi-empirical rule that simply states: $\lambda _{\sigma}\sim 2\lambda
_{\pi}$. Indeed, the theoretical calculations and experimental data  for 3d$^5$
configuration in FeO$_{6}^{9-}$ octahedra  yield:  $\lambda _{\sigma}^2 \approx
3.5 \lambda _{\pi}^2$ and $\lambda _{\sigma}^2 \approx (2.5\div 6.0)\lambda
_{\pi}^2$, respectively. \cite{Freund,Licht} The latter assumption concerns the
outgoing beyond the "local" approximation and seems to be more speculative.
Probably, this yields an overestimation for $t_{2u}(\pi)-e_g$ transition, but
underestimation for $t_{1u}(\pi)-e_g$ and $t_{1u}(\sigma)-t_{2g}$ transitions,
which intensity can be enhanced due to $t_{1u}(\pi)-t_{1u}(\sigma)$
hybridization. The more detailed quantitative description of the weak
dipole-allowed CT transitions needs the substantial expansion of our model
approach and further theoretical studies.

The calculated model contributions  of the dipole-allowed CT transitions to the
imaginary part of  dielectric function are presented in Figs.3a,b for
 conventional $Dq>0$ (Fig.3a) and unconventional $Dq<0$ (Fig.3b) sign
for the crystal field parameter. The top panels in both cases show the partial
contributions of different dipole-allowed transitions modeled by rather narrow
Lorentzians with linewidth $\Gamma = 0.5$ eV to clearly reveal the multiplet
structure. The lower panels present the overall contribution to the imaginary
part of dielectric function  of the dipole-allowed CT transitions. Here, the
Lorentzian linewidth is assumed to be  $\Gamma = 1.0$ eV for all contributions
to maximally reproduce the experimental situation.
 All the spectra are presented in the same  relative units.

As it was mentioned above, the energy spectrum of CT states in MnO$_{6}^{9-}$
octahedra does not depend on the sign of the crystal-field parameter $Dq$.
However, the intensities of certain CT transitions appear to be extremely
sensitive to the sign of $Dq$, or in other words, to the relative energy
position of $e_g$ and $t_{2g}$ orbitals.
 First of all, this concerns the
  $\pi -\pi$  and  $\sigma -\pi$  channels which manifest anomalously strong
  dependence on the $Dq$  sign with clearly seen spectral weight transfer
  from the composite band centered around $\sim 5$ eV given  $Dq>0$ (Fig.3a) to
 the composite band centered around $\sim 9$ eV given  $Dq<0$ (Fig.3b).
 From the other hand, the  $\sigma -\sigma$  and  $\pi -\sigma$
 channels which define the low-energy part of the overall CT band
 show no change with the $Dq$  sign inversion.
 Thus, the high-energy part of the overall CT band provides a very sensitive
 tool to examine the screening  effects for  crystal field  in the 3d$^5$
 configuration with the oxygen hole surrounding.

The theoretical findings are in reasonable quantitative agreement with
experimental spectra available. \cite{Okimoto} Indeed, a rather broad spectral
structure
 with distinctly revealed peaks near $5$ eV and $8\div 9$ eV  is observed
in the $\epsilon _{2}$ spectra for undoped LaMnO$_3$ \cite{Okimoto}
 (see inserts in Fig.3), although
only the band peaked near $5$ eV was assigned earlier to the O 2p-Mn 3d charge
transfer. The high-energy peak at $8\div 9$ eV was assigned by Okimoto {\it et
al}. to O 2p-La 5d interband transitions \cite{Okimoto} albeit this was argued
rather on the quantitative considerations than either calculations. However,
Arima and Tokura in their optical study of different perovskite-type RMO$_3$
(R=La, Y, M = Sc, Ti, V, Cr, Mn, Fe, Co, Ni, Cu) \cite{Arima} have shown that
similar band is commonly observed for all both LaMO$_3$ and YMO$_3$ compounds.
In other words, it seems more natural to ascribe this band rather to the
high-energy edge of the O 2p-M 3d CT transitions, than the O 2p-La 5d, O 2p-Y
4d. \cite{Arima} Comparing   experimental spectra \cite{Okimoto} with simulated
ones we may unambiguously assign the high-energy feature around $8\div 9$ eV to
the high-energy edge of the O 2p-Mn 3d CT transitions given the unconventional
sign  of the crystal field parameter $Dq < 0$ (see Fig.3b). In other words,
comparing the experimentally observed relative intensities of $5$ eV and $8\div
9$ eV bands, we may conclude that the oxygen hole in the CT states of
MnO$_{6}^{9-}$ center can give rise to the strong overscreening of the crystal
field parameter $Dq$ resulting in the sign inversion. Generally speaking, we
believe that this  conclusion needs  theoretical and experimental
substantiation  in the further studies of the CT states and transitions both in
LaMnO$_3$ and another 3d oxides.
 Nevertheless, one should note that irrespective of the numerical
value and sign of $Dq$, the low-energy spectral range of the overall CT band
consists of a series of transitions with increasing intensity beginning from
the lowest dipole-forbidden $t_{1g}(\pi)-e_g$  peaked at $1.7$ eV, weakly
dipole-allowed $t_{2u}(\pi)-e_g$ peaked at $2.5$ eV,
 relatively more intensive, but weak
dipole-allowed  $t_{1u}(\pi)-e_g$ peaked at $3.5$ eV, and, finally, strong
dipole-allowed  $t_{1u}(\sigma)-e_g$ transition peaked at $4.7$ eV,
respectively. It should be emphasized one more that the multi-band structure
with a wide spectral range of the order of $10$ eV represents one of the
characteristic features of  the O 2p - Mn 3d charge transfer optical response
in LaMnO$_3$ which has to be first taken into account before any theoretical
treatment of experimental spectra.

Interestingly, that with slight substitution in doped systems like
La$_{1-x}$Sr$_{x}$MnO$_3$ the weak low-energy edge band  at $1.7$ eV gradually
disappears \cite{Jung,Takenaka} with the simultaneous shift of the high-energy
bands to the lower frequencies. Our model allows to associate this effect with
the localization of the doped holes in the upper purely oxygen orbitals like
$t_{1g}(\pi)$, $t_{2u}(\pi)$. \cite{Avvakumov} Then the lowering of the
electron density for these states would result in the lowering of the intensity
for the appropriate CT bands. The red shift of the high-energy bands can result
from the screening effects induced by oxygen holes mainly for the Racah
parameter $A$.

Concluding this Section, we would like comment recent paper by N.N. Kovaleva
{\it et al.}. \cite{Kovaleva} In frames of the fully-ionic shell model the
authors explored the role of electronic and ionic polarization energies and
estimated the optical CT energies. To the best of our knowledge, this is first
attempt to give the realistic picture of different charge fluctuations in
manganites for which the polarization energies are crucial. In what concerns
the optical properties of LaMnO$_3$ the authors suggest that the band at $\sim
5$ eV is associated with the fundamental O 2p - Mn 3d CT transition, whereas
the band at $\sim 2$ eV is rather associated with the presence of Mn$^{4+}$
and/or O$^{-}$ self-trapped holes in  probably non-stoichiometric LaMnO$_3$
compound. Broad band peaked near $9$ eV is assigned to O 2p - La 5d CT
transition. However, their assignment is based on the numerical results
obtained in frames of the semi-empirical shell model with the full neglect of
the orbital degeneracy, many-electron intra-atomic correlations, and crystal
field effects both for manganese and oxygen electrons. On the other hand,
namely these effects are shown here to be responsible for multi-band structure
of the CT optical response. In a sense, the fundamental O 2p - Mn 3d CT energy
evaluated in Ref.\onlinecite{Kovaleva} to be $5.6$ eV represents a mean value
of the respective CT energies.

\section{Conclusions}
In  frames of a rather conventional quantum-chemical cluster approach, which
combines the crystal field and the ligand field models
 we have examined different CT states
and O 2p-Mn 3d CT transitions in MnO$_{6}^{9-}$ octahedra. The many-electron
dipole transition matrix elements were calculated using the Racah algebra for
the cubic point group. Simple "local" approximation allowed to calculate the
relative intensity for all dipole-allowed $\pi -\pi$ and $\sigma -\sigma$ CT
transitions. We present a self-consistent description of the CT bands in
insulating stoichiometric  LaMnO$_3$. Our analysis shows  the multi-band
structure of the CT optical response
  with the weak low-energy edge at $1.7$ eV, associated with forbidden
$t_{1g}(\pi)-e_{g}$ transition and a series of the high-energy weak and
strong dipole-allowed high-energy transitions
starting from $2.5$ and $4.5$ eV,
 respectively, and   extending up to nearly $11$ eV. The most intensive features
 are associated with
 two strong composite bands near $4.6\div 4.7$ eV
and $8\div 9$ eV, respectively,  resulting from the superposition of the dipole-allowed
  $\sigma -\sigma$ and $\pi -\pi$ CT transitions.
 These theoretical findings
 are in quantitative agreement with experimental
spectra available. We examined the effects of the sign of the crystal field
parameter $Dq$ and showed that the $\pi -\pi$  and  $\sigma -\pi$  channels
contrary to $\sigma -\sigma$  and  $\pi -\sigma$ ones manifest anomalously
strong
  dependence on the $Dq$  sign with clearly seen spectral weight transfer
  from the composite band centered around $\sim 5$ eV given  $Dq>0$  to
 the composite band centered around $\sim 9$ eV given  $Dq<0$.
 Thus, the high-energy part of the overall CT band provides a very sensitive
 tool to examine the screening  effects for the crystal field  in the 3d$^5$
 configuration with the oxygen hole surrounding.
 The experimental data point to
a strong overscreening of the crystal-field parameter $Dq$
in the CT states of MnO$_{6}^{9-}$ centers.
In addition, we would like
 emphasize the specific role of the intra-atomic correlation effects. It seems,
 the actual spectral picture of the CT optical response
 is determined on equal footing both
 by  the intra-atomic d-d electron-electron repulsion  and
 single electron  effects.
 We believe that these and many other semi-quantitative conclusions drawn from our
model will stimulate the further theoretical and experimental studies of the CT
states and transitions both in LaMnO$_3$ and another  3d oxides.
 We did not consider the inter-center CT transitions. Its role in optical
 response remains so far unclear, however, appropriate intensities seem to be small
 because these are proportional to small inter-center d-d transfer integrals squared.

\section{Acknowledgments}
The discussions with N.N. Loshkareva, Yu.P. Sukhorukov, E.A. Ganshina,  V.S.
Vikhnin, R. Hayn, S.-L. Drechsler  are acknowledged.
 The research described in this publication was supported in part by grant SMWK
of the Ministry of Science and Art of Saxony. The author
would like to thank for  hospitality Institut f\"{u}r Festk\"{o}rper- und Werkstofforschung
Dresden, where part of this work was made.
The author acknowledges a partial support from the
 Award No.REC-005 of the U.S. Civilian Research \& Development Foundation for the
Independent States of the Former Soviet Union (CRDF), Russian Ministry
 of Education, grant E00-3.4-280, and
Russian Foundation for Basic Researches, grant 01-02-96404.

\begin{widetext}
\appendix*
\section{ Numerical values for  $6j$- and $6\Gamma$-coefficients that one needs
to calculate the dipole transition matrix elements}

$$
\left\{ \begin{array}{ccc}
3/2 & 1 & 5/2 \\
1/2 & 2 & 1/2
\end{array}
\right\} = -\frac{1}{\sqrt{15}}\, ;\,
\left\{ \begin{array}{ccc}
3/2 & 1 & 3/2 \\
1/2 & 2 & 1/2
\end{array}
\right\} = \frac{1}{2\sqrt{10}}\, ;\,
\left\{ \begin{array}{ccc}
3/2 & 0 & 3/2 \\
1/2 & 2 & 1/2
\end{array}
\right\} = \frac{1}{2\sqrt{2}}\, ;\,
$$
$$
\left\{ \begin{array}{ccc}
A_{2} & E & E \\
E & A_{1}  & A_{2}
\end{array}
\right\} =
\left\{ \begin{array}{ccc}
A_{2} & E & E \\
E & A_{2}  & A_{1}
\end{array}
\right\} = - \frac{1}{\sqrt{2}}\, ;\,
\left\{ \begin{array}{ccc}
A_{2} & E & E \\
E & E  & E
\end{array}
\right\} =  \frac{1}{2}\, ;\,
$$
$$
-\left\{ \begin{array}{ccc}
T_{1} & T_{1} & E \\
E & A_{1}  & T_{1}
\end{array}
\right\} =
\left\{ \begin{array}{ccc}
T_{1} & T_{1} & E \\
E & A_{2}  & T_{2}
\end{array}
\right\} =
\left\{ \begin{array}{ccc}
T_{2} & A_{2} & T_{1} \\
E & T_{1}  &  E
\end{array}
\right\} =
\left\{ \begin{array}{ccc}
T_{2} & A_{2} & T_{1} \\
E & T_{2}  &  E
\end{array}
\right\} = - \frac{1}{\sqrt{6}}\, ;\,
$$
$$
\left\{ \begin{array}{ccc}
T_{2} & T_{1} & T_{2} \\
E & T_{1}  &  T_{2}
\end{array}
\right\} =
\left\{ \begin{array}{ccc}
T_{1} & T_{1} & T_{2} \\
E & T_{2}  &  T_{1}
\end{array}
\right\} =
-\left\{ \begin{array}{ccc}
T_{2} & T_{1} & T_{2} \\
E & T_{2}  &  T_{1}
\end{array}
\right\} =
-\left\{ \begin{array}{ccc}
T_{1} & T_{1} & T_{2} \\
E & T_{1}  &  T_{2}
\end{array}
\right\} =\frac{1}{6} \, ;\,
$$
$$
\left\{ \begin{array}{ccc}
T_{1} & T_{1} & E \\
E & E  &  T_{1}
\end{array}
\right\} =
\left\{ \begin{array}{ccc}
T_{1} & T_{1} & E \\
E & E  &  T_{2}
\end{array}
\right\} =
\left\{ \begin{array}{ccc}
T_{1} & T_{1} & T_{2} \\
E & T_{2}  &  T_{2}
\end{array}
\right\} =
$$
$$
-\left\{ \begin{array}{ccc}
T_{1} & T_{1} & T_{2} \\
E & T_{1}  &  T_{1}
\end{array}
\right\} =
-\left\{ \begin{array}{ccc}
T_{2} & T_{1} & T_{2} \\
E & T_{1}  &  T_{1}
\end{array}
\right\} =
-\left\{ \begin{array}{ccc}
T_{2} & T_{1} & T_{2} \\
E & T_{2}  &  T_{2}
\end{array}
\right\} =\frac{1}{2\sqrt{3}}\, .
$$
\end{widetext}

%
%
%
%
%
%

\end{document}